\documentclass[prb, twocolumn]{revtex4-1}
\usepackage{amsmath}
\usepackage{amsfonts}
\usepackage[pdftex]{graphicx}
\usepackage[utf8]{inputenc}

\begin{document}

\title{A Bayesian interpretation of abrupt phase transitions}

\author{Sergio Davis}
\homepage{http://www.gnm.cl/~sdavis}
\email{sdavis@gnm.cl}

\author{Joaqu\'{\i}n Peralta}
\email{jperalta@gnm.cl}

\author{Yasm\'{\i}n Navarrete}
\email{yasmin@gnm.cl}

\author{Diego Gonz\'alez}
\email{dgonzalez@gnm.cl}
 
\author{Gonzalo Guti\'errez}
\email{gonzalo@fisica.ciencias.uchile.cl}

\affiliation{Grupo de Nanomateriales, Departamento de F\'{i}sica, Facultad de Ciencias,
Universidad de Chile, Casilla 653, Santiago, Chile}

\date{\today}

\begin{abstract}
The formalism used in describing the thermodynamics of abrupt (or first-order)
phase transitions is reviewed as an application of maximum entropy inference. In 
this treatment, we show that the concepts of transition temperature, latent heat and 
entropy difference between phases will inevitably have an equivalent in any
problem of inferring the result of a yes/no question, given information in the 
form of expectation values.
\end{abstract}



\maketitle

\section{Introduction}

Jaynes' proposal of the principle of maximum entropy (PME) as a general tool of
probabilistic inference~\cite{Jaynes1957,Jaynes2003} is remarkable in that it is both widely 
used~\cite{Presse2013} and somewhat controversial~\cite{Uffink1995,CardosoDias1981}. 
It asserts that the most unbiased probability distribution $P$ given some fixed knowledge 
$\mathcal{I}$ is the one that maximizes Shannon's information entropy,

\begin{equation}
S[P(x|\mathcal{I})] = -\sum_x P(x|\mathcal{I})\log_2 P(x|\mathcal{I})
\end{equation}
while being consistent with said knowledge. This is because $S[P]$ is a measure
of uncertainty~\cite{Shannon1948} or lack of knowledge about the degrees of
freedom (represented collectively by $x$ in the notation above) and, maximizing it leads 
to the probabilistic model containing the least amount of information, but nevertheless 
able to reproduce the features one demands of it. As this is a process of
inference it cannot be deductive: predictions derived from the maximum entropy
model may be proved wrong by subsequent measurements, and this reflects an
incompleteness of the fixed knowledge used to constrain the maximization. 

Jaynes' interpretation of the formalism of statistical mechanics sees it as just
the application of this principle of maximum entropy, valid in all statistical
inference, to the case of a macroscopic number of particles (and degrees of
freedom). In this situation the predictions are almost perfectly
sharp, with uncertainties vanishing as $1/\sqrt{N}$, with $N$ the number of
degrees of freedom. This is all well described for the case of thermodynamic
equilibrium of a single phase. However, how this information-theoretical interpretation
manifests itself in the case of the study of phase transitions, and what can we
learn from this, is an issue which has not been so extensively clarified. For
instance, in his book on probability theory~\cite{Jaynes2003} (p. 602), Jaynes 
wrote in a somewhat cryptic footnote, that

\begin{quote}
``... in statistical mechanics the relative probability $P_j/P_k$ of two different
phases, such as liquid and solid, is the ratio of their partition functions
$Z_j/Z_k$, which are the normalization constants for the sub-problems of
prediction within one phase. In Bayesian analysis, the data are indifferent
between two models when their normalization constants become equal; in
statistical mechanics the temperature of a phase transition is the one at which
the two partition functions become equal...''
\end{quote}

This suggests that the problem of liquid-solid phase transition, or in fact,
any phase transition, can be posed as a model comparison problem,
and therefore the transition temperature and the free energy can be given an
information-theoretical meaning. In this work, in order to remove all
particularities of thermodynamics from the treatment of abrupt (or first-order)
phase transitions, we present a parallel of the formalism used in
first-order phase transitions based entirely on the application of the PME. We
introduce a simple game, the ``disc throwing'' game, and answer two questions
related to it by means of the PME. In the answers to these questions we will recover 
the concepts of transition temperature, Helmholtz free energy and the rule that
imposes its equality for the two phases at the coexistence point.

The rest of the paper is organized as follows. In section II we review the main
features of the maximum entropy formalism. Section III shows an illustration of
PME inference, while section IV describes and solves the disc throwing game
problem. In section V we expose the perfect parallel between the solution of this 
problem and that of the coexistence of two phases in thermodynamical equilibrium. 
Finally we conclude with some remarks.

\section{Maximum Entropy Inference}

Consider a system (in the most general sense of the word) having $N$ discrete degrees of freedom $\vec{x}$
and being fully described in statistical terms by a function $f(\vec{x})$ with
known expectation value $f_0$. Knowledge of $f_0$ is symbolically represented by $\mathcal{I}$. 
According to the PME, the most unbiased model is the one that maximizes the Gibbs-Shannon entropy functional

\begin{equation}
S = -\sum_{\vec{x}} P(\vec{x}|\mathcal{I})\log_2 P(\vec{x}|\mathcal{I})
\label{eq_shannon}
\end{equation}
subject to the constraint $\mathcal{I}$, i.e., to 

\begin{equation}
\Big<f(\vec{x})\Big> = f_0.
\end{equation}

Maximization under this constraint, and the always implicit constraint of proper
normalization of the probability, is achieved by the inclusion of Lagrange
multipliers $\lambda$ and $\mu$ respectively, after which the problem reduces to the 
maximization of the augmented function

\begin{eqnarray}
\tilde{S} = -\sum_{\vec{x}} P(\vec{x}|\mathcal{I})\log_2 P(\vec{x}|\mathcal{I}) +  \nonumber \\ 
\lambda (f0 - \sum_{\vec{x}} P(\vec{x}|\mathcal{I})f(\vec{x})) + \mu (1 - \sum_{\vec{x}} P(\vec{x}|\mathcal{I})).
\end{eqnarray}

This leads to the well-known maximum entropy (MaxEnt) model 

\begin{equation}
P(\vec{x}|\lambda) = \frac{1}{Z(\lambda)}\exp(-\lambda f(\vec{x}))
\label{eq_maxent}
\end{equation}
in which we have changed the notation from the purely abstract $P(\vec
x|\mathcal{I})$ to the mode concrete $P(\vec x|\lambda)$, given that the
parameter $\lambda$ distinguishes between all the possible states of knowledge
compatible with the possible values of $f_0$. The function $Z$,

\begin{equation}
Z(\lambda) = \sum_{\vec{x}} \exp(-\lambda f(\vec{x})).
\end{equation}
is known as the partition function. The Lagrange multiplier $\lambda$ is 
usually determined as the unique solution of 

\begin{equation}
-\frac{\partial}{\partial \lambda}\ln Z(\lambda) = f_0.
\label{eq_DlnZ}
\end{equation}

If the degrees of freedom contained in $\vec x$ are continuous, Shannon entropy
needs to be replaced with the relative entropy 

\begin{equation}
S = -\int d\vec{x} P(\vec{x}|\mathcal{I}\wedge I_0)\log_2 \frac{P(\vec{x}|\mathcal{I}\wedge I_0)}{P(\vec{x}|I_0)}
\end{equation}
where $I_0$ denotes an ``initial'' state of knowledge. The solution to the
maximum entropy problem is now

\begin{equation}
P(\vec{x}|\mathcal{I}\wedge I_0) = \frac{1}{Z(\lambda)}P(\vec{x}|I_0)\exp(-\lambda f(\vec{x}))
\label{eq_maxent_sol}
\end{equation}
with 
\begin{equation}
Z(\lambda) = \int d\vec{x} P(\vec{x}|I_0) \exp(-\lambda f(\vec{x})).
\end{equation}

In both cases (discrete and continuous degrees of freedom), the maximized
entropy has a value 

\begin{equation}
S = \ln Z(\lambda) + \lambda f_0.
\label{eq_max_S}
\end{equation}

Now, we have just described the formalism of the canonical ensemble if we think of the system as
composed by $n$ particles with position $\vec{r}_i$ and momentum $\vec{p}_i$
(with $i$=1,...,$n$) and the descriptor function as the Hamiltonian $f=\mathcal{H}(\vec{r}_1,\ldots,\vec{r}_n,\vec{p}_1,\ldots,\vec{p}_n)$.
Then Eq. \ref{eq_maxent} is the canonical distribution where we identify $\lambda=\beta=1/(k_B T)$. 

In thermodynamic notation, Eq. \ref{eq_max_S} reads,

\begin{equation}
S(\beta)/k_B = \ln Z(\beta) + \beta E(\beta)
\label{eq_max_S2}
\end{equation}

If we introduce the Helmholtz free energy $\beta F(\beta)=-\ln Z(\beta)$, Eq.
\ref{eq_max_S2} reduces to

\begin{equation}
S(\beta)/k_B = \beta (E(\beta)-F(\beta))
\end{equation}
i.e., using $\beta=1/k_B T$,
\begin{equation}
F(T) = E(T)-T S(T).
\end{equation}

\section{An illustration of the maximum entropy formalism}

Suppose we have a swimming pool full of plastic balls (all spherical) of
different radii. The average volume of a ball is $V$. What is the average radius?

We have the constraint,

\begin{equation}
\left<\frac{4}{3}\pi r^3\right> = V,
\label{eq_vconstr}
\end{equation} 
which is equivalent to 

\begin{equation}
\left<r^3\right> = \frac{3}{4\pi}V,
\end{equation}
from which the most unbiased model for $r$ is 

\begin{equation}
P(r|\lambda) = \frac{1}{Z(\lambda)}\exp(-\lambda r^3)\Theta(r).
\label{eq_sol_radii}
\end{equation}

The partition function is given by

\begin{equation}
Z(\lambda) = \int_0^\infty dr \exp(-\lambda r^3) = \Gamma(4/3)\lambda^{-1/3},
\end{equation}
therefore, the value of $\lambda$ is determined from 

\begin{equation}
-\frac{\partial}{\partial \lambda}\ln Z(\lambda) = \frac{1}{3\lambda} =
\frac{3}{4\pi}V,
\end{equation}
i.e., $\lambda=4\pi/(9V)$. Note that from inspection of Eq. \ref{eq_sol_radii}
and the fact that $\lambda$ is positive, the most probable radius is zero and the probability 
monotonically decreases with $r$. The expectation of $r$ is then

\begin{widetext}
\begin{eqnarray}
\left<r\right> = \frac{1}{Z(\lambda)}\int_0^\infty dr r\exp(-\lambda r^3) =
\frac{3^{1/3}}{3}\frac{\Gamma(2/3)}{\Gamma(4/3)}\cdot\left(\frac{3V}{4\pi}\right)^{1/3} \approx 0.729011 \sqrt[3]{\left<r^3\right>}.
\end{eqnarray}
\end{widetext}

From this example we learn two things. First, the Lagrange multiplier $\lambda$ is larger
for small $V$, and this is expected given that the smaller $V$ is, the possible radii
are more concentrated around zero and therefore there is less uncertainty about
the value of the radius. This means the constraint of known $V$ (Eq.
\ref{eq_vconstr}) has greater ``weight'' for smaller $V$. Second, the expected radius is less than the na\"{\i}ve
estimate $r_0=\sqrt[3]{\left<r^3\right>}$, valid in the case where all the balls
have the same radius.
As the distribution function $P(r|\lambda)$ decreases from $r=0$ onward, there are more balls 
with $r \leq r_0$ than with $r > r_0$ and thus the estimate $\left<r\right>$ is skewed towards zero.

\section{A simple disc throwing game}

\begin{figure}[h]
\begin{center}
\includegraphics[scale=0.3]{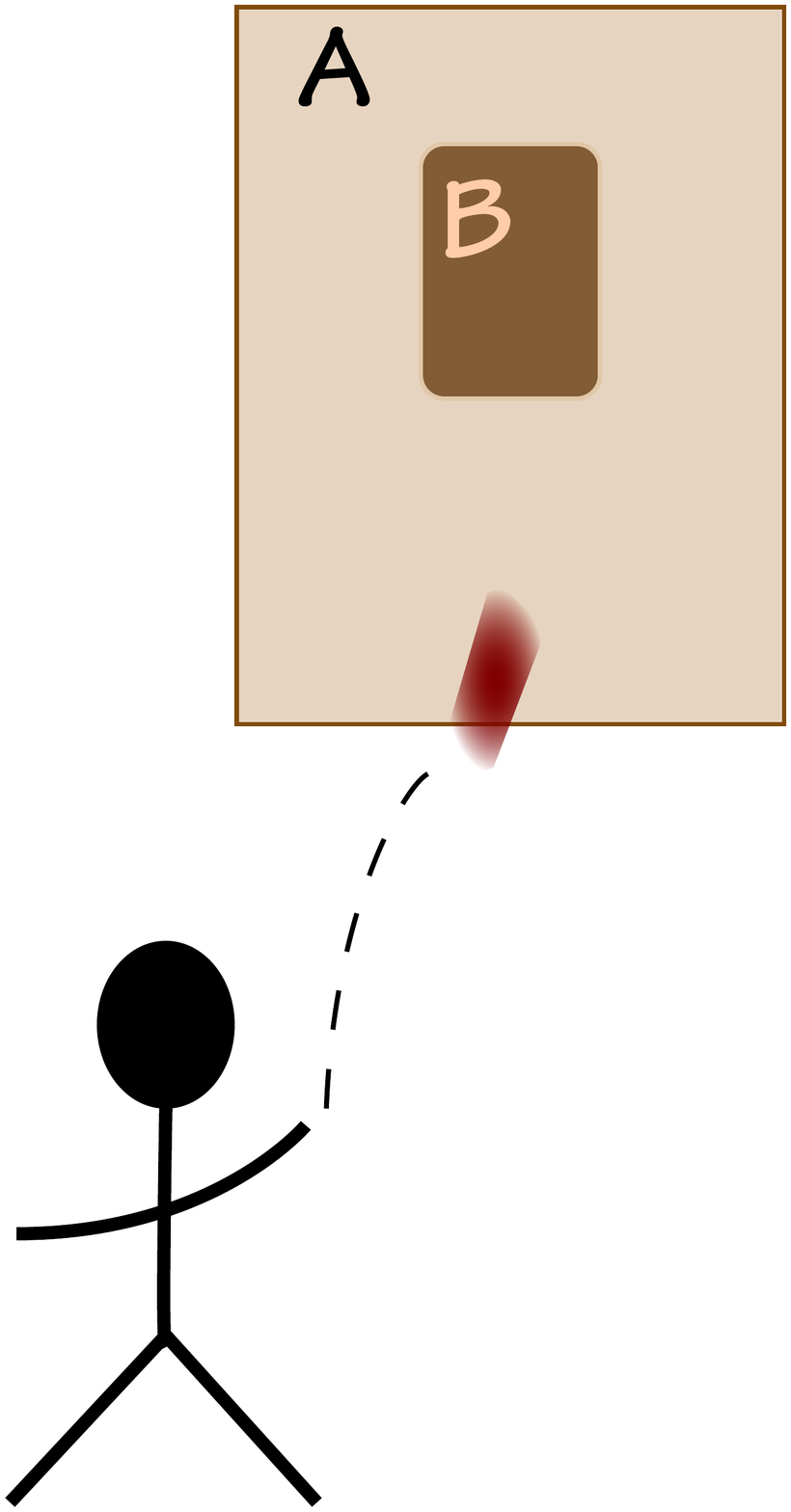}
\end{center}
\caption{Schematic representation of the disc throwing game.}
\label{fig_game}
\end{figure}

Suppose a player can throw a disc into a surface $A$ (with area $\Sigma_A$),
containing within it a smaller surface $B$ (with area $\Sigma_B < \Sigma_A$).
We consider $A$ and $B$ to be disjoint regions, as shown in Fig. \ref{fig_game}. A
successful hit within $B$ gives the player $n_B$ points, whereas a hit inside
$A$ (outside $B$) gives $n_A$ points to the player (as hitting $B$ is
more difficult, $n_B > n_A$). This is similar to the game ``rayuela'' as is known in
some South American countries.

We can present two questions about this game:

\begin{enumerate}
\item[(a)] With only the information laid out above, and particularly without
knowing anything about the performance of the player, what probability should
one assign to hitting $B$? 
\item[(b)] Now consider the player has obtained an average score of
$\overline{n}$ in the past (over enough trials to be considered a reliable
average). What probability should one assign now to hitting $B$?
\end{enumerate}

In (a) the intuitive answer is that the probabilities of hitting either $A$ or
$B$ are completely determined by their areas. In fact, considering each landing
point as a coordinate inside $A$, and because such points are mutually
exclusive, exhaustive alternatives and there is symmetry under exchange, we can easily 
see that

\begin{equation}
\frac{P(A|\mathcal{I}_1)}{P(B|\mathcal{I}_1)} = \frac{\Sigma_A}{\Sigma_B}
\end{equation}

From this, given that landing in $A$ or $B$ constitute mutually exclusive and
exhaustive propositions, $P(A|\mathcal{I}_1)+P(B|\mathcal{I}_1)=1$. Therefore,

\begin{equation}
P(\alpha|\mathcal{I}_1) = \frac{\Sigma_\alpha}{\Sigma_A+\Sigma_B}
\end{equation}
with $\alpha=A,B$. The predicted score of the player, with just the information
we have in (a), is then

\begin{equation}
\overline{n} = \frac{\Sigma_A n_A + \Sigma_B n_B}{\Sigma_A+\Sigma_B}.
\label{eq_score_prior}
\end{equation}

We see that probabilities are governed only by the ratio $\Sigma_A/\Sigma_B$,
and we can conclude that always $P(A|\mathcal{I}_1) > P(B|\mathcal{I}_1)$, given that the area of $B$ is 
considerably smaller. Now, what happens in (b) is that we have to constrain the 
inference to this new information, given in the form of an expectation value. 
We invoke the law of large numbers and assume $\big<n\big>=\overline{n}$, then 
the most unbiased probability for either result given $\overline{n}$, according to 
the PME, is (using Eq. \ref{eq_maxent_sol}), 

\begin{equation}
P(\alpha|\mathcal{I}_2) = \frac{1}{Z(\lambda)}\Sigma_\alpha \exp(-\lambda n_\alpha)
\end{equation}
with 
\begin{equation}
Z(\lambda) = \Sigma_A \exp(-\lambda n_A)+\Sigma_B \exp(-\lambda n_B),
\label{eq_Z}
\end{equation}
and 
\begin{equation}
-\frac{\partial}{\partial \lambda}\ln Z(\lambda) = \overline{n}.
\label{eq_dlnZ_new}
\end{equation}

After explicitly using the result of Eq. \ref{eq_Z} in Eq. \ref{eq_dlnZ_new} and
some algebra, we have that 

\begin{equation}
\Sigma_A \left(\overline{n} - n_A\right)\exp(-\lambda n_A) = \Sigma_B \left(n_B - \overline{n}\right)\exp(-\lambda n_B)
\end{equation}
from which it follows that $\lambda$ is given by

\begin{equation}
\lambda(\overline{n}) = -\frac{1}{n_B-n_A}\left[\ln \Sigma_A-\ln \Sigma_B + \ln \frac{\overline{n}-n_A}{n_B-\overline{n}}\right].
\label{eq_lambda_general}
\end{equation}

In order to simplify notation, let us introduce 

\begin{eqnarray}
\Delta n = n_B - n_A, \\
\Delta S = S_B - S_A = \ln \Sigma_B - \ln \Sigma_A.
\end{eqnarray}

Then Eq. \ref{eq_lambda_general} reads,

\begin{equation}
\lambda \Delta n - \Delta S = \ln \frac{n_B-\overline{n}}{\overline{n}-n_A}
\label{eq_lambda_compact}
\end{equation}

It is clear that, when $\lambda=0$, Eq. \ref{eq_lambda_compact} implies 

\begin{equation}
\frac{n_B-\overline{n}}{\overline{n}-n_A} = \frac{\Sigma_A}{\Sigma_B}.
\end{equation}
which is nothing but the result of section (a), Eq. \ref{eq_score_prior}. This
happens when the reported average score $\overline{n}$ is the same as predicted
from the area information alone. This reflects a complete lack of ability from the 
player to control the hitting spot, because the results do not differ from pure
``random'' shots. However, if $\overline{n}$ is not consistent with Eq.
\ref{eq_score_prior}, then $\lambda \neq 0$ and the ratio between probabilities is not 
simply the ratio of the respective areas, but it is given by

\begin{equation}
\frac{P(A|\mathcal{I}_2)}{P(B|\mathcal{I}_2)} = \frac{\Sigma_A}{\Sigma_B} \exp(-\lambda (n_A-n_B))
\label{eq_ratio_general}
\end{equation}
i.e., defining $\Delta \ln P=\ln P(B|\mathcal{I}_2)-\ln P(A|\mathcal{I}_2)$, 

\begin{equation}
\Delta \ln P = \Delta S - \lambda \Delta n,
\label{eq_ratio_general_compact}
\end{equation}
or, if we define $F_\alpha = n_\alpha-S_\alpha/\lambda$, we have 

\begin{equation}
\Delta \ln P = -\lambda \Delta F. 
\label{eq_delta_F}
\end{equation}

Therefore, the most probable outcome ($A$ or $B$) would be the one with lowest
value of $F$.

After comparing Eqs. \ref{eq_ratio_general_compact} and \ref{eq_lambda_compact}, 
the ratio of probabilities is given by

\begin{equation}
\frac{P(A|\mathcal{I}_2)}{P(B|\mathcal{I}_2)} = \frac{n_B-\overline{n}}{\overline{n}-n_A}.
\end{equation}

There will be an interesting value of $\overline{n}$, namely the average
$(n_A+n_B)/2$, where $P(A|\mathcal{I}_2)=P(B|\mathcal{I}_2)$. In this case we are maximally
uncertain with respect to which region the player will hit, i.e., we have
``canceled out'' all the information we had from the areas by using the average
score. This situation corresponds to a ``critical value'' of the Lagrange multiplier,

\begin{equation}
\lambda_0 = \lambda\Big(\frac{n_A+n_B}{2}\Big) = \frac{\Delta S}{\Delta n}.
\end{equation}

\section{Bayesian Thermodynamics}

Perhaps it will be striking to the reader (at first) to notice that we have
replicated the formalism used to study first-order phase transitions in
thermodynamical systems. Imagine the two regions $A$ and $B$ of the game
introduced previously, as regions in phase space corresponding, for instance, to 
liquid and solid, respectively. We can relate the area of each region $\Sigma$ to the 
volume in phase space occupied by each of the thermodynamic phases, and in this sense, the quantity 

\begin{equation}
S=\ln \Sigma
\end{equation}
is readily interpreted as the Boltzmann entropy (taking $k_B$=1). 
Therefore the most probable phase (i.e., the most stable phase in thermodynamical terms) is, in absence of any other
information, the one with the largest value of entropy. This is the same
situation as in the microcanonical ensemble~\cite{Callen1985}. 

When we have information about the expected (or average) score $\overline{n}$, 
analogous to the measured internal energy $E$ of a thermodynamical system ($n_A$ and $n_B$
are then the internal energies for the liquid and solid phases, respectively), 
what decides the most probable phase is, according to Eq. \ref{eq_delta_F}, the difference 
in the quantity

\begin{equation}
F = n - S/\lambda
\end{equation}
which is precisely the Helmholtz free energy (under the identification
$\lambda=\beta=1/T$),

\begin{equation}
F = E - TS.
\end{equation}

If we are given a low enough value of energy (close to the energy of the ideal
solid) then, despite the fact that the liquid phase has a larger entropy, we are 
forced to conclude that the system is in one of the (relatively) few solid phase points. 
Because this reversal of our prediction after knowing $\overline{n}$ is strikingly
unexpected, this situation is described by a large value of the Lagrange multiplier 
$\lambda$ which, in the context of thermodynamics, corresponds to a low value of
temperature $T$.

The limiting situation when we cannot claim to know the most probable
phase happens when $\Delta F=0$, which is the condition of thermodynamic phase coexistence.
The Lagrange multiplier then is $\lambda_0=\Delta S/\Delta n$, or, in
thermodynamic notation,

\begin{equation}
T_0=L/\Delta S(T_0),
\label{eq_latentheat}
\end{equation}
where $L$ is the latent heat associated with the first-order phase transition
and $\Delta S(T_0)$ is the entropy difference at the transition temperature $T_0$.

All these equivalences are summed up in Table \ref{tbl_equiv}.

\begin{table}
\begin{tabular}{|c|c|}
\hline
Throwing Game & Thermodynamics \\
\hline
Logarithm of area ($\ln \Sigma)$ & Entropy ($S$) \\
Game average score ($\overline{n})$ & Internal energy ($E$)\\
Score difference ($\Delta n$) & Latent heat ($L$) \\
Critical multiplier ($1/\lambda_0$) & Transition temperature ($T_0$) \\
\hline
\end{tabular}
\caption{Equivalences between concepts arising in the analysis of the throwing
game and thermodynamical concepts.}
\label{tbl_equiv}
\end{table}

\section{``Thermodynamics'' of a binary question}

We have shown that the concepts of latent heat and critical temperature apply to
the case of the disc throwing game. However, one may ask, how general are these
results? What are the conditions the game must fulfill for these concepts to be
applicable?

In the most general terms, consider any question $Q$ which can be answered in
the affirmative/negative, and two different states of knowledge: the prior state
$\mathcal{I}_0$ and a state $\mathcal{I}_1=\mathcal{D}\wedge \mathcal{I}_0$ which includes a new piece of
knowledge (or datum) $\mathcal{D}$.

From Bayes' theorem we have

\begin{eqnarray}
P(Q|\mathcal{I}_1) = \frac{P(Q|\mathcal{I}_0)P(\mathcal{D}|Q \wedge {I}_0)}{P(\mathcal{D}|\mathcal{I}_0)} \\
P(\neg Q|\mathcal{I}_1) = \frac{P(\neg Q|\mathcal{I}_0)P(\mathcal{D}|\neg Q \wedge {I}_0)}{P(\mathcal{D}|\mathcal{I}_0)}
\end{eqnarray}

Dividing both equations and cancelling the common denominator
$P(\mathcal{D}|\mathcal{I}_0)$ we have 

\begin{equation}
\frac{P(Q|\mathcal{I}_1)}{P(\neg Q|\mathcal{I}_1)} =
\frac{P(Q|\mathcal{I}_0)}{P(\neg Q|\mathcal{I}_0)} \frac{P(\mathcal{D}|Q\wedge
\mathcal{I}_0)}{P(\mathcal{D}|\neg Q\wedge \mathcal{I}_0)},
\end{equation}
which can be written in logarithmic form as

\begin{equation}
(\Delta \ln P)_{\mathcal{I}_1} = (\Delta \ln P)_{\mathcal{I}_0} +
\mathcal{E}(\mathcal{D}).
\label{eq_lnP_general}
\end{equation}
 
Note that

\begin{equation}
\mathcal{E}(\mathcal{D}) = \ln P(\mathcal{D}|Q\wedge \mathcal{I}_0)-\ln P(\mathcal{D}|\neg
Q\wedge \mathcal{I}_0)
\end{equation}
is the only quantity dependent on the datum $\mathcal{D}$, and thus encapsulates
the effect this datum has on the balance between $Q$ and $\neg Q$. To understand the meaning of this quantity, let us define, as 
in Jaynes~\cite{Jaynes2003} (p. 91) the evidence $e(Q|\mathcal{D})=\ln P(Q|\mathcal{D})-\ln P(\neg
Q|\mathcal{D})$ in favor or against $Q$ (versus $\neg Q$). Then Eq. \ref{eq_lnP_general} reads,

\begin{equation}
\mathcal{E}(\mathcal{D}) = e(Q|\mathcal{D}\wedge \mathcal{I}_0) - e(Q|\mathcal{I}_0) = (\Delta e)_\mathcal{D}.
\end{equation}
and therefore $\mathcal{E}$ is precisely the change in evidence when
incorporating the datum $\mathcal{D}$. Comparing with Eq.
\ref{eq_ratio_general_compact} written as 

\begin{equation}
(\Delta \ln P)_{\mathcal{I}_2} = (\Delta \ln P)_{\mathcal{I}_1} -\lambda \Delta n
\end{equation}
we see that, for the disc throwing game, the evidence brought in by the datum
$\overline{n}$ is $\mathcal{E}(\overline{n})=-\lambda \Delta n$. This
corresponds, \emph{mutatis mutandi}, to $\mathcal{E}(\overline{E})=-L/T$ for a
first-order phase transition, therefore

\begin{equation}
\frac{T}{L} = -\frac{1}{\mathcal{E}(\overline{E})}
\end{equation}

Thus in this latter context, temperature $T$ (in units of $L$) 
measures the effect, in terms of evidence relevant to which phase the system is,
brought in by including the value of energy $\overline{E}$ in the inference
procedure. A large value of the evidence translates into temperatures close to
absolute zero, favoring the low entropy phase.

\section{Concluding remarks}

We have shown that, because to every yes/no question we can associate a change
in evidence introduced by a new fact, there exist analogous quantities to the free energy
difference between phases and the transition temperature, that are closely connected 
to this change in evidence. When the evidence is strong enough to completely
cancel out our initial judgements about the probability of one phase over another and leave us
undecided, the ``weight'' of this evidence is proportional to the transition inverse temperature.

Thus, in this view, the problem of thermodynamic equilibrium between phases is
seen as answering the question: \emph{is the system in phase A if we know that
its average energy is $\overline{E}$?} in a Bayesian/maximum entropy formalism. 
The concepts of transition temperature and free energy arise naturally as consequences 
of this inference framework, and therefore are not intrinsic properties of the systems or the phases.

\section{Acknowledgements}

SD and JP gratefully acknowledges partial funding from FONDECYT grant 1140514.

\bibliography{phasetrans}
\bibliographystyle{apsrev}

\end{document}